\documentclass[aps,prl,twocolumn,showpacs,preprintnumbers,superscriptaddress,amsmath]{revtex4}
\usepackage{txfonts}
\usepackage{amssymb}
\usepackage{graphicx}
\usepackage{dcolumn}
\usepackage{bm}

\begin{document}

\title{Transmission of doughnut light through a bull's eye structure}
\author{Lu-Lu Wang}
\author{Xi-Feng Ren\footnote{renxf@ustc.edu.cn}}
\author{Rui Yang}
\address{Key Laboratory of Quantum Information, University of Science and Technology of China, Hefei
230026, People's Republic of China}
\author{Guang-Can Guo}
\author{Guo-Ping Guo\footnote{gpguo@ustc.edu.cn}}
\address{Key Laboratory of Quantum Information, University of Science and Technology of China, Hefei
230026, People's Republic of China}
\begin{abstract}
We experimentally investigate the extraordinary optical transmission
of doughnut light through a bull's eye structure. Since the
intensity is vanished in the center of the beam, almost all the
energy reaches the circular corrugations (not on the hole), and
excites surface plasmons which propagate through the hole and
reradiate photons. The transmitted energy is about 32 times of the
energy input on the hole area. It is also interesting that the
transmitted light has a similar spatial shape with the input light
even though the diameter of the hole is much smaller than the
wavelength of light.

\end{abstract}
\pacs{ 78.66.Bz,73.20.MF, 71.36.+c}

\maketitle

The phenomenon of extraordinary optical transmission (EOT) through
metallic films which were perforated by nanohole arrays was first
observed a decade ago\cite{Ebbesen98}. It is generally believed that
surface plasmons (SPs) in metal surface play a crucial role in this
process, during which photons first transform into SPs and then back
to photons again\cite{Moreno,liu}. Such SPs are involved in a wide
range of applications\cite{Barnes03,Ozbay06}. The report of EOT
phenomenon attracts considerable attention because it shows that
more light than Bethe's prediction could be transmitted through the
holes\cite{Bethe}. This stimulates much fundamental research and
promotes subwavelength apertures as a core element of new optical
devices. For EOT in periodic hole arrays, not only the polarization
properties\cite{Elli04,Koer04,RenAPL} but also the spatial mode
properties\cite{ren06,ren062} are widely discussed. Even for a
single aperture surrounded by circular corrugations, we can also get
high transmission efficiencies and a well-defined spectrum since the
periodic corrugations act as an antenna to couple the incident light
into SPs\cite{thio,Lezec}.

Usually, the light transmitted through the subwavelength holes can
be divided into two parts: one is the directly transmitted light and
the other comes from the surface plasmon assisted transmission
process. Here we present a new method to eliminate the influence of
the first part in EOT phenomenon by using a doughnut input light and
a bull's eye structure. Since the intensity is null in the center of
the beam, there is no light illuminating on the single hole
directly. Almost all the energy reaches the circular corrugations,
and excites SPs which propagate through the hole and reradiate
photons (as shown in Fig.1 ). It is also interesting that the
transmitted light has a similar spatial shape with the input light
even though the diameter of the hole is much smaller than the
wavelength of light.

\begin{figure}[b]
\includegraphics[width=8.0cm]{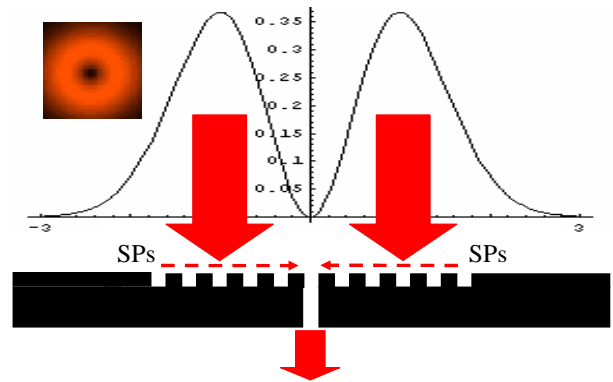}
\caption{Sketch map of our protocol. The typical doughnut light has
an intensity null on the beam axis. Almost all the energy reaches
the circular corrugations, excite surface plasmons which propagate
through the hole and reradiate photons.}
\end{figure}

\begin{figure}[b]
\includegraphics[width=8.0cm]{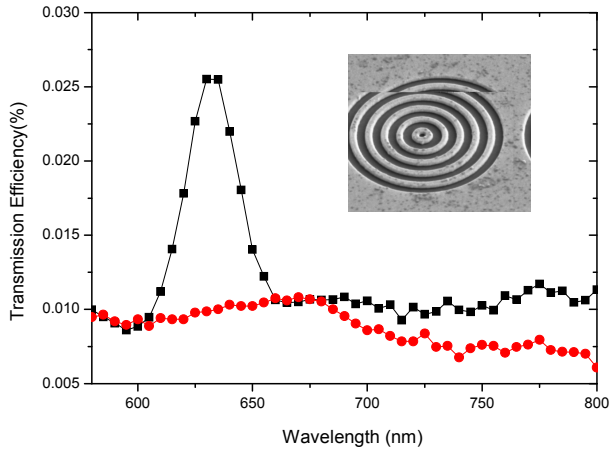}
\caption{(Color online) Transmission efficiency as a function of
wavelength for bull's eye structure(Black square dots) and similar
structure without hole in center(Red round dots). Inset is a
scanning electron microscope picture of our bull's eye
structure(groove periodicity, 500 nm; groove depth, 60 nm; hole
diameter, 250nm; film thickness, 135 nm).}
\end{figure}

Inset of Fig. 2 is a scanning electron microscope picture of our
bull's eye structure. The thickness of the gold layer is $135$ $nm$.
The cylindrical hole($250$ $nm$ diameter) and the grooves are
produced by a Focused Ion Beam Etching system (FIB, DB235 of FEB
Co.). The grooves have a period of $500 nm$ with the depth $60 nm$
and width $250 nm$. Transmission spectra of the hole array are
recorded by a Silicon avalanche photodiode (APD) single photon
detector coupled with a monochromator through a fiber. White light
from a stabilized tungsten-halogen source passes through a single
mode fiber and a polarizer (only vertical polarized light can pass),
then illuminates on the sample. The hole array is set between two
lenses with the focus of $35mm$. The light exiting from the hole
array is launched into the monochromator. The transmission spectra
are shown in Fig. 2(Black square dots), in which the transmission
efficiency is determined by normalizing the intensity of transmitted
light over the intensity before the sample. At the resonant
frequency (632.8 nm in the experiment), the transmission efficiency
is about $2.55\%$, much higher than that of the non-resonant case.
To verify the phenomenon does not come from the direct transmitted
light, we use another sample as a comparison. The new sample also
has a Bull's eye geometry, but without hole in center. The
transmission efficiencies are all about $1.0\%$ and there is no
transmission peak as shown in Fig. 2(Red round dots), which verify
that the transmission peak for bull's eye structure come from the
surface plasmons assisted transmission process. In the following
experiments, we use the bull's eye structure with a hole in center
to investigate the extraordinary optical transmission of doughnut
light.

The typical doughnut light is produced by changing its orbital
angular momentum (OAM), which is associated with the transverse
phase front of a light beam. Light field of photons with OAM can be
described by means of Laguerre-Gaussian ($LG_p^l$) modes with two
mode indices $p$ and $l$\cite{Allen92}. The $p$ index gives the
number of radial nodes and the $l$ index represents the number of
the $2\pi$-phase shifts along a closed path around the beam center.
Light with an azimuthal phase dependence $e^{-il\varphi}$ carries a
well-defined OAM of $l\hbar$ per photon\cite{Allen92}. When $l=0$,
the light is in the general Gaussian mode, while when $l\neq 0$, the
associated phase discontinuity produces an intensity null on the
beam axis. If the mode function is not a pure LG mode, each photon
of this light is in a superposition state, with the weights dictated
by the contributions of the comprised different $l$th angular
harmonics. For the sake of simplification, we can consider only LG
modes with the index $p=0$. Computer generated holograms
(CGHs)\cite{ArltJMO,VaziriJOB}, a kind of transmission holograms,
are used to change the winding number $l$ of LG mode light. Inset of
Fig. 3. shows part of a typical CGH($n=1$) with a fork in the
center. Corresponding to the diffraction order $m$, the $n$ fork
hologram can change the winding number $l$ of the input beam by
$\Delta l_m=m*n$. In our experiment, we use the first order
diffraction light ($m=1$) and the efficiency of our CGHs is about
$30\%$. Superposition mode is produced using a displaced
hologram\cite{VaziriJOB}, which is particularly suitable for
producing superposition states of $LG_0^l$ mode with the Gaussian
mode beam.

\begin{figure}[b]
\includegraphics[width=8.0cm]{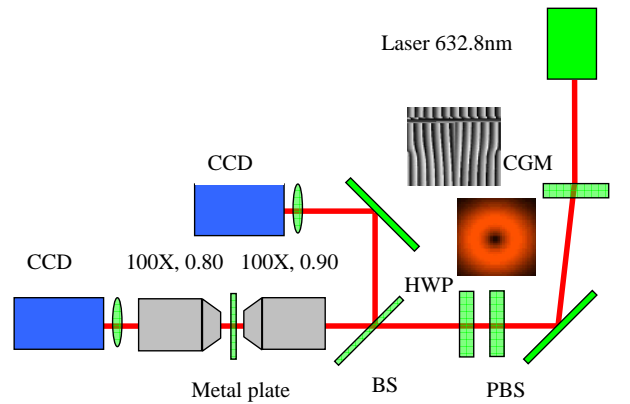}
\caption{Experimental setup. A computer generated hologram(CGH) is
used to change the OAM of the laser beam. The polarized laser beam
is directed into the microscope and focused on the metal plate using
a 100X objective lens (Nikon, NA=0.90). Transmitted light is
collected by another 100X objective lens (Nikon, NA=0.80). Inset,
pictures of part of a typical CGH($n=1 $) and produced light with
the first order mode.}
\end{figure}

The experimental setup is shown in Fig. 3. The OAM of the laser
light(632.8nm wavelength) is changed by a CGH, while the
polarization is controlled by a polarization beam splitter (PBS,
working wavelength 632.8 nm) followed by a half wave plate (HWP,
working wavelength 632.8 nm). The polarized laser beam is directed
into the microscope and focused on the metal plate using a 100X
objective lens (Nikon, NA=0.90) with a diameter about $3.8\mu m$.
The CCD camera before the objective lens is used to adjust the
position of the hole structure. Transmitted light is collected by
another 100X objective lens (Nikon, NA=0.80) and recorded by another
CCD camera. The relative position of the beam center to the hole is
estimated as follows: We detect the transmission of the Gaussian
beam with the sample moved by a three-dimensional stage (Suruga
Seiki Co., Ltd. B71-80A). When the center of beam is coincided with
the center of hole, the maximum transmission is achieved.  A CCD
camera is also used as an assistant to observe the picture directly.
Since the doughnut light is produced via the movement of hologram
which does not affect the optical path, we can realize the protocol
that the position of zero electric field coincides with the center
of the hole.

\begin{figure}[b]
\includegraphics[width=8.0cm]{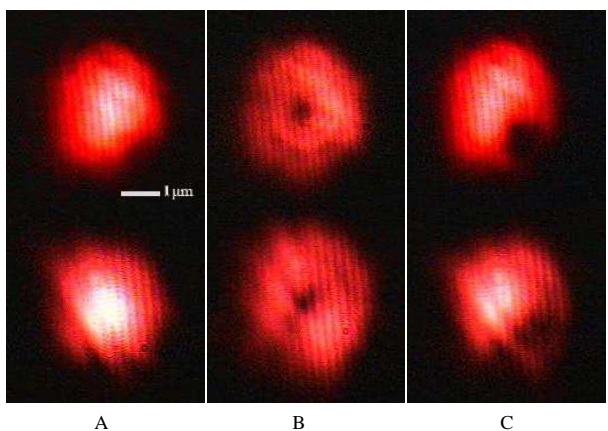}
\caption{CCD pictures of light beam before (upper) and after (lower)
the bull's eye structure. The light power is decreased to give clear
pictures. A, B, C are the cases for light with Gaussian mode
($l=0$), the first order mode($l=1$) and a typical superposition
mode $(a\left| 0\right\rangle +b\left| 1\right\rangle
)/\sqrt{a^2+b^2}$ (where $a$ and $b$ are real numbers)
respectively.}
\end{figure}
Transmission efficiencies are measured for light with Gaussian mode
($l=0$), the first order mode($l=1$) and a typical superposition
mode $(a\left| 0\right\rangle +b\left| 1\right\rangle
)/\sqrt{a^2+b^2}$, where $a$ and $b$ are real numbers). When the
hologram is placed in the beam center, the OAM of the first
diffraction order light is $1$, while for hologram in the beam edge,
the OAM is $0$. In the middle part, the output light is in the
superposition mode of $0$ and $1$. The results for $0$ and $1$ order
mode light are $2.55\%$, and $2.28\%$ respectively. The transmission
efficiency for the superposition mode light is between the upper two
cases and can be changed with the ratio of $a$ and $b$ when we move
the hologram. In all the cases, transmission efficiency is much
larger than the value obtained from the classical
theory\cite{Bethe}. The reason is that the interaction of the
incident light and surface plasmon is made allowed by coupling
through the grating momentum and obeys conservation of the momentum
\begin{equation}
\overrightarrow{k}_{sp}=\overrightarrow{k}_{0}\pm
i\overrightarrow{G}_{x}\pm j\overrightarrow{G}_{y},
\end{equation}
where $\overrightarrow{k}_{sp}$ is the surface plasmon wave vector,
$\overrightarrow{k}_{0}$ is the component of the incident wave
vector that lies in the plane of the sample,
$\overrightarrow{G}_{x}$ and $\overrightarrow{G}_{y}$ are the
reciprocal lattice vectors, and i, j are integers.
$\overrightarrow{G}_{x,y}=2\pi/d_{x,y}$ are the lattice vectors in
the $x,y$ directions respectively, and $d_{x,y}$ are the periodicity
of the structure in the $x$ and $y$ direction. While in the
practical experiments, the Eq.1 can not be satisfied simply, because
many parameters can influence the resonant frequency, for example,
the thickness of the metal film, the width of the grooves, as
mentioned in \cite{Degi}. Due to the symmetry of the Bull's eye
structure, the polarization of the light has no influence on the
whole process. We can see that the transmission efficiency for
Gaussian mode light is larger than that of the first order mode
light. Although it is hard to give a precise explanation, the
possible factors may be that the additional transmissions of
Gaussian mode light from directly passing light, SPs excited from
the hole edge by scattering, and lower propagating loss in the hole.
This lower loss comes from the waveguide property of the hole in
which the Gaussian mode light has a higher transmission efficiency
than that of other modes as shown in\cite{Moreno,ruan}.

Calculation shows that the energy in the beam center (250nm
diameter) is only about $0.04\%$ of the whole doughnut light.
Comparing with the SPs assisted transmission efficiency $1.28\%$, we
can find that the transmitted energy is about 32 times of the
directly illuminating light on the hole area. This can be the
evidence that the transmitted light in the case of doughnut mode
results from the surface plasmon assisted transmission process.

CCD pictures are also taken for the three cases as shown in Fig. 4.
The light power is decreased to give clear pictures. It is
interesting that the spatial shape of the light was still preserved
after the plasmon assisted transmission process, even though the
hole diameter (250 nm) is much smaller than the light wavelength
(632.8 nm). Since the spatial shape of the light is determined by
its OAM, which is associated with the transverse phase front of a
light beam, we can conclude that the OAM of the photons are not
influenced in this process. It has been proven in many works that
the phase of the photons can be preserved in the surface plasmons
assisted transmission process, here we show that the helical
wavefront of photons can also be transferred to SPs and carried by
them\cite{ren06}.

In conclusion, we investigate the extraordinary optical transmission
phenomenon through a subwavelength aperture surrounded by circular
corrugations when the light is in the doughnut shape. Since all the
energy reaches the circular corrugations but not on the hole, the
directly transmitted light can be ignored. The present experiment
could provide intriguing prospects for both the exploiting of the
surface plasmon based devices and the study of fundamental physics
issues.

This work was funded by the National Basic Research Programme of
China (Grants No.2009CB929600 and No. 2006CB921900), the Innovation
funds from Chinese Academy of Sciences, and the National Natural
Science Foundation of China (Grants No. 10604052 and No.10874163).


\begin{references}
\bibitem{Ebbesen98}  T.W. Ebbesen, H. J. Lezec, H. F. Ghaemi, T. Thio, and P. A. Wolff,  Nature(London) 391, 667 (1998).

\bibitem{Moreno}  L. Martin-Moreno, F. J. Garcia-Vidal, H. J. Lezec, K. M. Pellerin, T. Thio, J. B. Pendry, and T. W. Ebbesen, Phys. Rev. Lett. 86, 1114
(2001).
\bibitem{liu} Haitao Liu, Philippe Lalanne, Nature(London) 452, 728
(2008).
\bibitem{Barnes03} W. L. Barnes, A. Dereux, T. W. Ebbesen,  Nature 424, 824 (2003).
\bibitem{Ozbay06} E. Ozaby, Science 311, 189 (2006).
\bibitem{Bethe}  H. A. Bethe, Phys. Rev. 66, 163 (1944).
\bibitem{Elli04} J. Elliott, I. I. Smolyaninov, N. I. Zheludev, and A. V. Zayats, Opt. Lett. 29, 1414 (2004).

\bibitem{Koer04} K. J. K. Koerkamp, S. Enoch, F. B. Segerink, N. F. van Hulst, and L. Kuipers, Phys. Rev. Lett. 92, 183901 (2004).
\bibitem{RenAPL} X. F. Ren, G. P. Guo, Y. F. Huang, Z. W. Wang, and G.
C. Guo, Appl. Phys. Lett. 90, 161112 (2007).
\bibitem{ren06} X. F. Ren, G. P. Guo, Y. F. Huang, Z. W. Wang, and G. C. Guo, Opt.
Lett. 31, 2792, (2006).
\bibitem{ren062} X. F. Ren, G. P. Guo, Y. F. Huang, C. F. Li, and G. C. Guo, Europhys. Lett. 76, 753 (2006).

\bibitem{thio}T. Thio, K. M. Pellerin, R. A. Linke, H. J. Lezec, and T.
W. Ebbesen, Opt. Lett. 26, 1972 (2001).
\bibitem{Lezec} H. J. Lezec, A. Degiron, E. Devaux, R. A. Linke,
L. Martin-Moreno, F. J. Garcia-Vidal, T. W. Ebbesen, Science 297,
820 (2002).
\bibitem{Allen92}  L. Allen, M. W. Beijersbergen, R. J. C. Spreeuw, and J.
P. Woerdman, Phys.Rev.A.  45, 8185, (1992).

\bibitem{ArltJMO} J. Arlt, K. Dholokia, L. Allen, and M. Padgett. J. Mod. Opt.
45, 1231 (1998).

\bibitem{VaziriJOB} A. Vaziri, G. Weihs, and A. Zeilinger. J. Opt. B:
Quantum Semiclass. Opt 4, s47 (2002).

\bibitem{Degi} A. Degiron and T.W. Ebbesen, Opt. Express 12, 3694
(2004).

\bibitem{ruan} Zhichao Ruan, and Min Qiu, Phys. Rev. Lett. 96, 233901 (2006).
\end{references}
\end{document}